\documentclass{llncs}
\usepackage{graphicx}
\usepackage{comment}
\begin{document}
\sloppy
\title{Respondent-Driven Sampling in Online Social
  Networks\thanks{This material is based upon work supported by the
    National Science Foundation under Grant No. 1111016, and through
    Grant No. K23MH079215 from the National Institute of Mental
    Health, National Institutes of Health. The final publication is available at link.springer.com. }}
\author{Christopher M. Homan\inst{1} \and Vincent Silenzio\inst{2}
  \and Randall Sell\inst{3}}
\institute{Rochester Institute of Technology, Rochester, NY. Email: \email{cmh@cs.rit.edu}
\and University of Rochester Medical Center, Rochester, NY. Email: \email{vincent\_silenzio@URMC.Rochester.edu}
\and Drexel University, Philadelphia, PA. Email:
\email{rls82@drexel.edu}}
\maketitle
\begin{abstract}
Respondent-driven sampling
(RDS) is a commonly used method for acquiring
data on hidden communities, i.e., those that lack
unbiased sampling frames or face social
stigmas that make their members unwilling to identify themselves.
Obtaining accurate statistical data about such communities is important because,
for instance, they often have different health burdens from the
greater population, and without good statistics it is hard and expensive to effectively reach them for
prevention or treatment interventions. Online social networks (OSN) have the
potential to transform RDS for the better. We present a new RDS
recruitment protocol for (OSNs) and show via simulation that it
outperforms the standard RDS protocol in terms of sampling accuracy and approaches the accuracy
of Markov chain Monte Carlo random walks.
\end{abstract}
\section{Introduction}
Respondent-driven sampling
(RDS)~\cite{hec:j:rdsone,salganik2004sampling,hec:u:rds,hec:j:rds-theory-08,wejnert_heckathorn_2008}
is a commonly used method to survey such communities as IV drug users, men who have 
sex with men, and sex workers~\cite{malekinejad2008using}; jazz
musicians~\cite{heckathorn2001finding}; unregulated
workers~\cite{bernhardt2012all}; native American subcommunities~\cite{walters2002health}; and other
hidden communities. RDS is a variant of snowball
sampling~\cite{tho:b:sampling} that uses a clever
recruitment protocol that: (1) helps ensure the confidentiality
of respondents and the anonymity of the target community and (2)
generates a relatively large number of recruitment waves, which
hypothetically leads to unbiased sampling estimators.

Unfortunately, in terms of sampling accuracy there is still a large gap between theory and
practice~\cite{wejnert_2009,gil-han:j:assess,tomas2011effect,goel_salganik_2010}. A
small body of
work~\cite{salganik2004sampling,hec:j:rds-theory-08,salganik2004sampling,hec:u:rds,hec:j:rds-theory-08,wejnert_heckathorn_2008},
most of which focuses on improving
the estimators on which RDS depends, deals with 
closing that gap.

This paper describes a new approach: leveraging the features of online
social networks (OSNs) to improve the sampling design. We
believe OSNs have the potential to dramatically transform RDS by enabling
better neighborhood recall, randomized and confidential recruitment, and other
improvements that allow it to better meet the assumptions on which the
estimators rest. Here we focus on one particular
modification, which is
based on the network that a recruitment
protocol generates, i.e., the network consisting of all respondents as
actors and having directed
ties between each respondent and those whom the respondent recruits.
The estimators for RDS typically assume that these so-called
\emph{recruitment networks} are 
arbitrary, although in practice they are essentially trees. Gile and
Handcock show~\cite{gil-han:j:assess} in simulation that this
discrepancy is a major source of the poor
performance they observe in established RDS estimators. 

Our main contribution is a new protocol where the 
recruitment networks are directed acyclic
graphs (DAGs). This protocol, while likely infeasible in many other settings,
seems well suited for RDS over OSNs. Using the same simulation-based experimental framework that
Gile and
Handcock~\cite{gil-han:j:assess} and Tomas and Handcock~\cite{tomas2011effect}
developed in their rather
comprehensive  assessments of RDS, we show that this new
protocol dramatically outperforms the standard RDS protocol and
approaches the sampling
accuracy of a Markov chain Monte
Carlo (MCMC) random walk (a process that typically satisfies standard RDS sampling assumptions).
It even outperforms a recruitment protocol that, superficially at least,
more closely resembles MCMC
walks than does ours. 

Our work is related to that of Gjoka et al.~\cite{Gjoka2011}, who
use the established RDS estimators to compare the performance of several different 
methods for passively---without the active participation of its users---crawling Facebook, including MCMC random
walks and breadth-first search.
By contrast, we are concerned primarily with methods that, due to 
confidentially concerns, require the active participation of those
sampled, and this leads different sampling dynamics.

In another closely related study, Wejnert and Heckathorn develop a
tool for conducting RDS over the World-Wide Web
they call WebRDS~\cite{wejnert_heckathorn_2008}. Their
system explicitly fixes the recruitment graph to be a tree. We, on the
other hand, study what happens precisely when we relax this
constraint.

\section{A Brief Overview of Respondent-Driven Sampling}
Heckathorn introduced RDS as a sampling protocol paired with an estimator~\cite{hec:j:rdsone}. The
protocol begins
with a small number of seed respondents from the target community,
who may be recruited in any fashion. Each respondent takes a survey,
and is then given a small number of recruitment coupons (e.g., three) to
distribute among other
members of the target community, each of which allows
whomever redeems it to take the survey (assuming that he or she meets the inclusion 
criteria). Each respondent is paid for taking the survey and for each
of the redeemed coupons he or she distributed. The process
continues until a target number of either recruitment waves or samples is
reached. 
Thus, RDS uses the social network of the hidden population itself to
do the work of subject identification, and in this regard it has been very
successful in finding hidden communities.  Couponing ensures the confidentiality of all
those surveyed, which is often a crucial concern for the communities RDS is designed to reach.

Though the recruitment protocol has remained stable, 
the estimators have evolved significantly over time as questions are
raised about each successive generation of estimators. We present here
what is known as the Volz-Heckathorn (VH)
estimator~\cite{hec:j:rds-theory-08}. Although probably not as widely
used as an earlier estimator due to Salganik and
Heckathorn~\cite{salganik2004sampling}, it is newer and has
been the subject of recent
papers~\cite{gil-han:j:assess,tomas2011effect,Gjoka2011} that
experimentally test its performance. 
In particular, Handcock and Gile
show that the VH estimator frequently outperforms the Salganik-Heckathorn estimator~\cite{gil-han:j:assess}.
 The assumptions underlying the VH estimator are:

\begin{enumerate}
\item \label{cond:graph} The network is connected and
  aperiodic.
\item \label{cond:one} Each respondent recruits exactly one person into the survey.
\item Each respondent chooses whom to recruit uniformly at random from
  all network relationships.
\item All relationships are reciprocal.
\item \label{cond:replace} Respondents are sampled with replacement (i.e., may be
  rerecruited into the survey).
\item \label{cond:recall} Respondents can accurately recall the number of people in the target community
  that they know. 
\end{enumerate}

It is fairly clear that in practice these assumptions, except possibly
the first one, never hold. In this paper,
we are particularly interested in
assumption~\ref{cond:replace}. In typical RDS settings most people lack
the time to respond more than once, since doing so often involves
travel, so this assumption
fails. Consequently, recruitment networks tend to look like trees.

It is worth noting that prior estimators rested on even stronger 
assumptions~\cite{hec:j:rdsone,salganik2004sampling}.
More recently, Handcock and Gile~\cite{handcock2010modeling} 
proposed newer estimators that depend on fewer
assumptions and that seem in their experiments to outperform earlier
estimators~\cite{gile2011network} (see
also~\cite{gile2011improved,tomas2011effect,gile2012diagnostics}).
Though their approach seems very promising, it is model based, and
such approaches themselves depend on assumptions that can be
difficult or impossible to validate.

Let $\{y_1,\ldots y_n\}$ be samples of some scalar property of a
networked population. Let each $d_i \in \{d_1, \ldots, d_n\}$ be the 
degree (number of network ties) of the person associated with each
sample. When the VH assumptions do hold, Markov chain Monte
Carlo (MCMC) theory suggests $\hat{y} = (\sum_{i=1}^n
  y_i/d_i)/(\sum_{i=1}^n 1/d_i)$ as an asymptotically unbiased estimator for the mean of
$\{y_1, \ldots,y_n\}$.

\section{Simulation-based experiments for assessing RDS}\label{sec:gh}

Gile and Handcock~\cite{gil-han:j:assess} and
Tomas and Gile~\cite{tomas2011effect}
 provide a pair of thorough critiques of the VH estimator. We 
adopted their methods to test our new recruitment protocol, so we
present them here in detail.

They simulate RDS over graphs drawn
randomly from an exponential random graph model (ERGM). In each experiment, $20\%$ of
the network nodes are labeled ``infected'' and the remaining are
``uninfected.'' The goal in these experiments is
to estimate the proportion of infected nodes in the population. Each
experiment fixes the ERGM and recruitment parameters, then repeats the
following steps 1000 times:
\begin{enumerate}
\item Generate a test graph from the ERGM.
\item Run an RDS simulation on the test graph; stop when 500 samples
  are made.
\item Estimate the proportion of infected nodes using VH.
\end{enumerate}

The ERGM parameters Gile and Handcock use are based on a CDC 
study~\cite{abdul2006effectiveness}. Network 
size ranges from $525$ to $1000$. They fix the expected degree at
seven. Expected \emph{activity ratio} is the mean degree of the
infected nodes divided by the mean degree of the uninfected nodes.
This ranges from one to three.  Expected homophily is defined here as
the expected number
of relationship between infected actors divided the expected number of
relationships between infected and uninfected actors. This ranges from
two to thirteen.

Seed nodes are drawn at random in proportion to their neighborhood
size, either from all nodes, just the infected nodes, or just the
non-infected nodes. The number of seeds ranges from 4 to 10.

For the recruitment parameters, each chosen node
recruits exactly two new nodes uniformly at random from its
``eligible'' network neighbors, where ``eligible'' is
either all neighbors (for sampling with replacement) or all neighbors
who have not yet been sampled (for sampling without replacement).
We call the without-replacement protocol ``RDS'' and the
with-replacement one ``REP.''  Note that RDS produces trees
as recruitment networks and REP produces arbitrary graphs.

 \section{A new DAG-based recruitment protocol}
\label{sec:new}
As Gile and Hancock show (see also
Fig.~\ref{fig:rdsrep-no-bias}--\ref{fig:rdsrep-non-burn}, which
reproduce in part their results), the RDS protocol, 
even with perfect randomness and response in the recruitment process,
results in significantly degraded performance under the VH estimator. 
But what if sampling with replacement were feasible? It seems
plausible do to so in an online setting, i.e., where the survey is
administered via the Web: if a respondent is recruited a
second time, all the respondent needs to do is log in to the website
where the survey is administered and the system can automatically
count the respondent's survey a second time (and send the respondent
additional electronic recruiting coupons) without requiring the
respondent to return to a physical polling site. 

The trickier part is in the recruitment dynamics. If we let 
respondents rerecruit freely, as in the REP protocol, then,  in order to
gain more money from survey incentives, they could collude to
rerecruit each other many more times than chance would predict, thus skewing the results. We propose to discourage
this behavior by allowing respondents to be rerecruited only if doing
so does not result in the recruitment graph containing a directed
cycle. The resulting recruitment graph is thus a directed acyclic graph. We call this protocol ``DAG.''

\section{Experiments and Results}
We use the same methods as Gile and Handcock, as we described in
section~\ref{sec:gh}. 
The major difference is that we consider two 
additional variants of the RDS
protocol: ``MCMC,'' in which each respondent recruits only one person
(with replacement),
chosen from that person's friend list uniformly at random, i.e., it is
a Markov chain Monte Carlo random walk and serves as a control case;
and ``DAG,'' as described in
Sect. \ref{sec:new}.
 
\begin{figure}
  \centering
  \includegraphics[scale=.70]{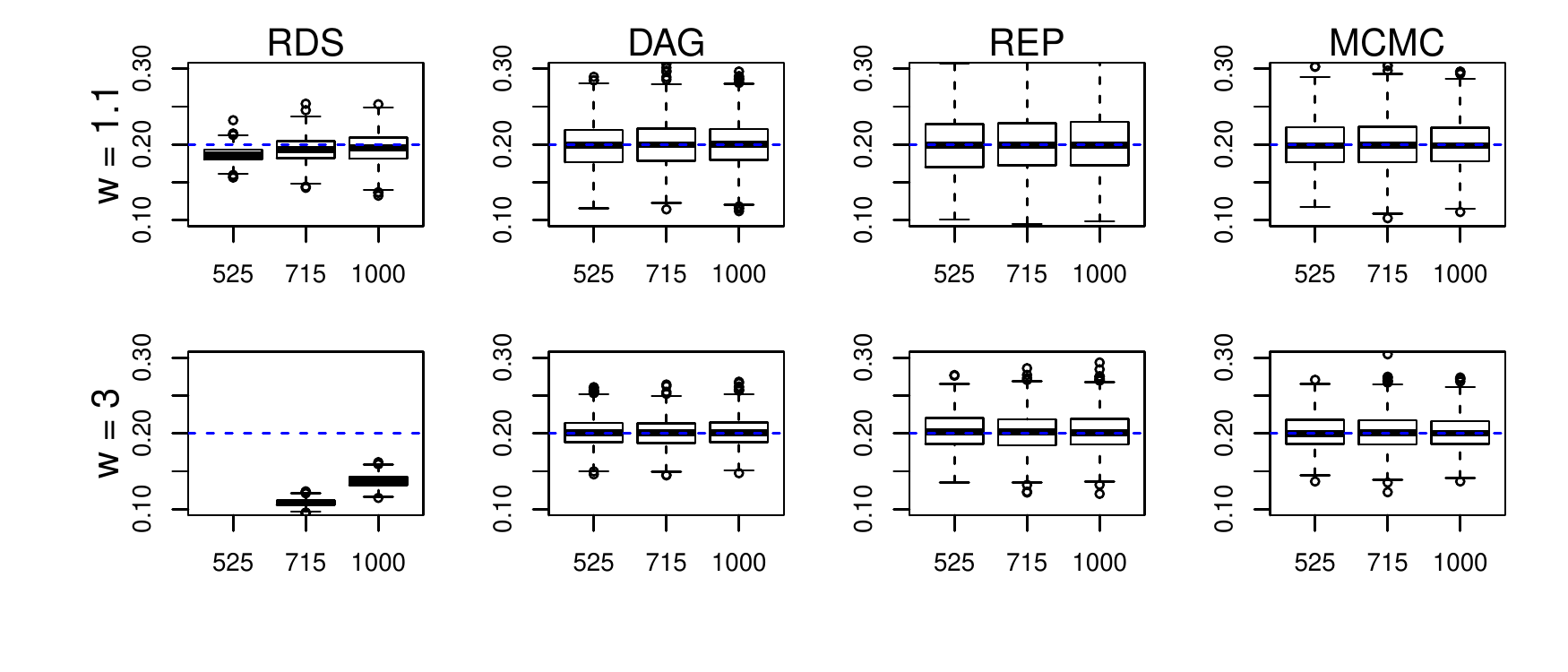}
\vspace{-1cm}
\caption{\label{fig:rdsrep-no-bias} Estimated size of infected population
  where seeds are drawn from the entire population with no burn-in.}
\end{figure}

\begin{figure}
  \centering
  \includegraphics[scale=.70]{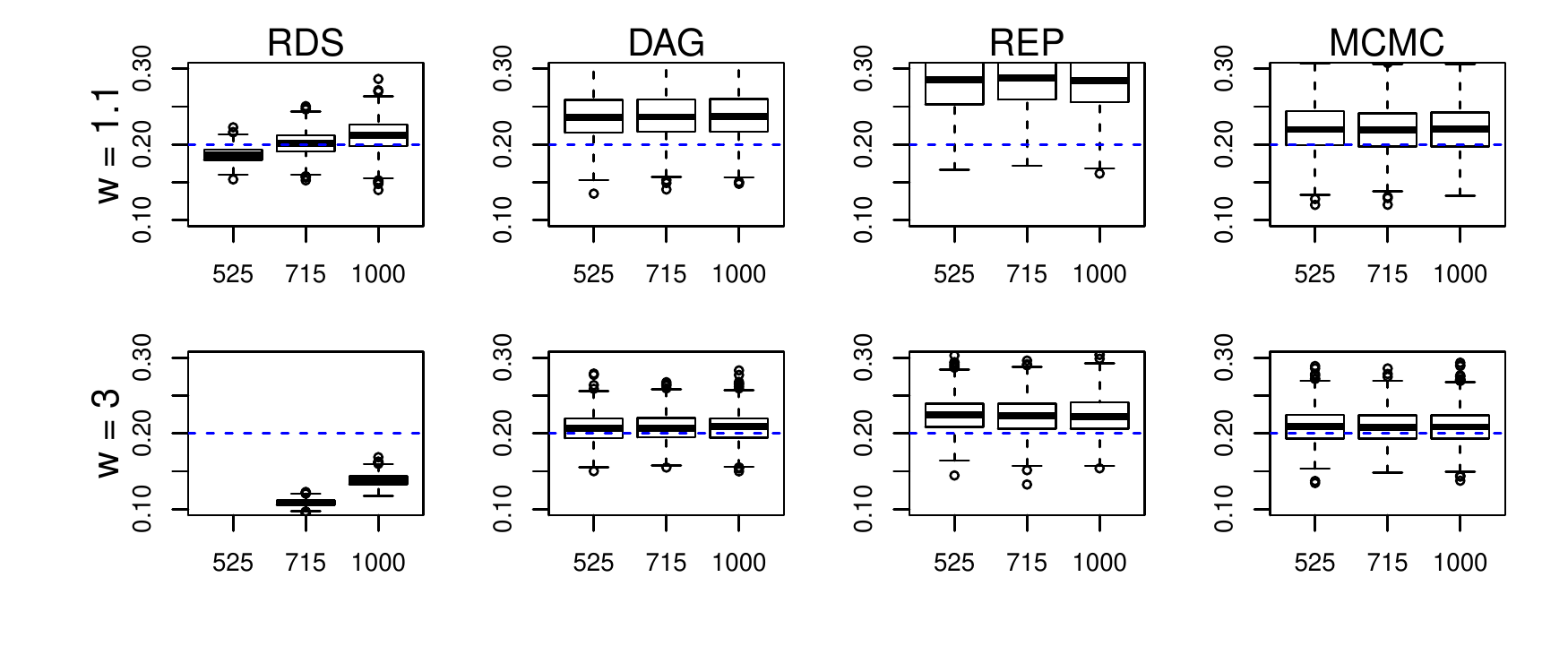}
\vspace{-1cm}
\caption{\label{fig:rdsrep-inf} Estimated size of infected population
  where seeds are drawn from the infected population only with no burn-in.}
\end{figure}

\begin{figure}
  \centering
  \includegraphics[scale=.70]{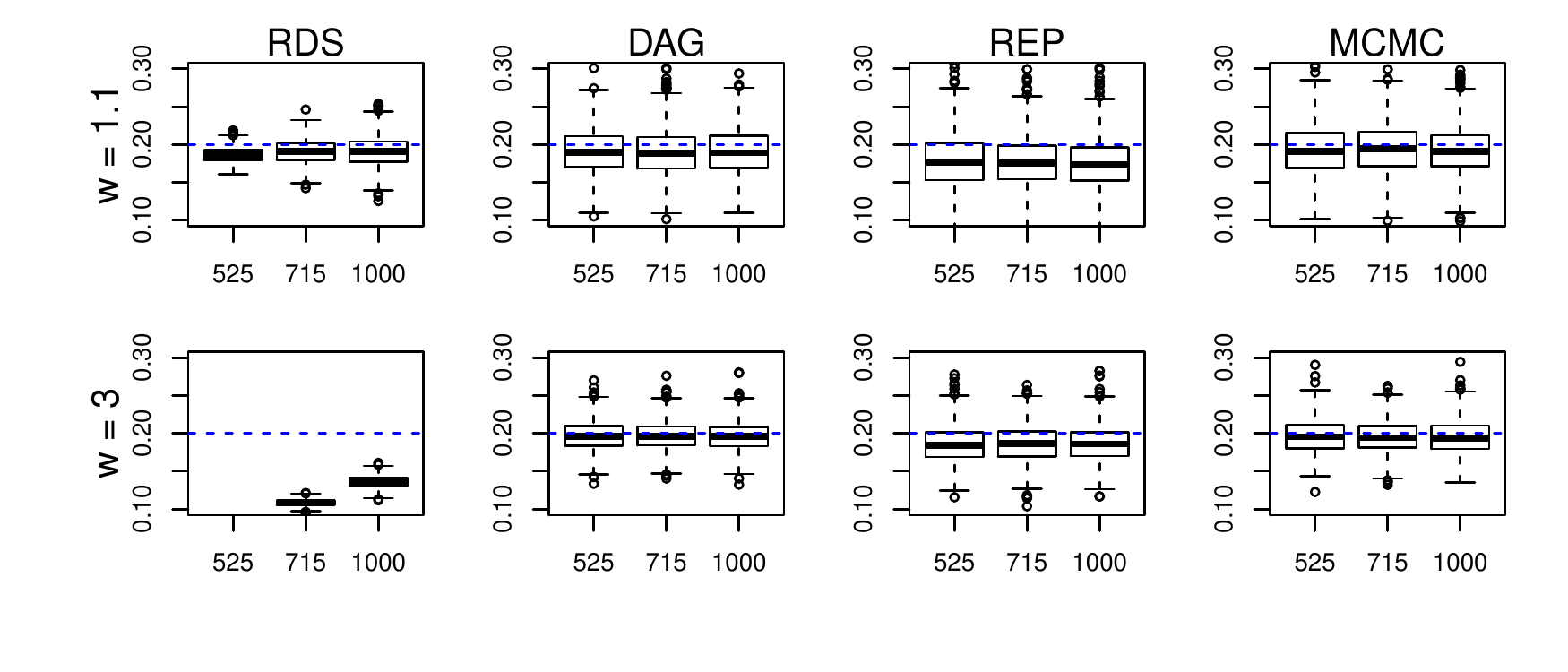}
\vspace{-1cm}
\caption{\label{fig:rdsrep-non} Estimated size of infected population
  where seeds are drawn from the noninfected population only with no burn-in.}
\end{figure}

\begin{figure}
  \centering
  \includegraphics[scale=.70]{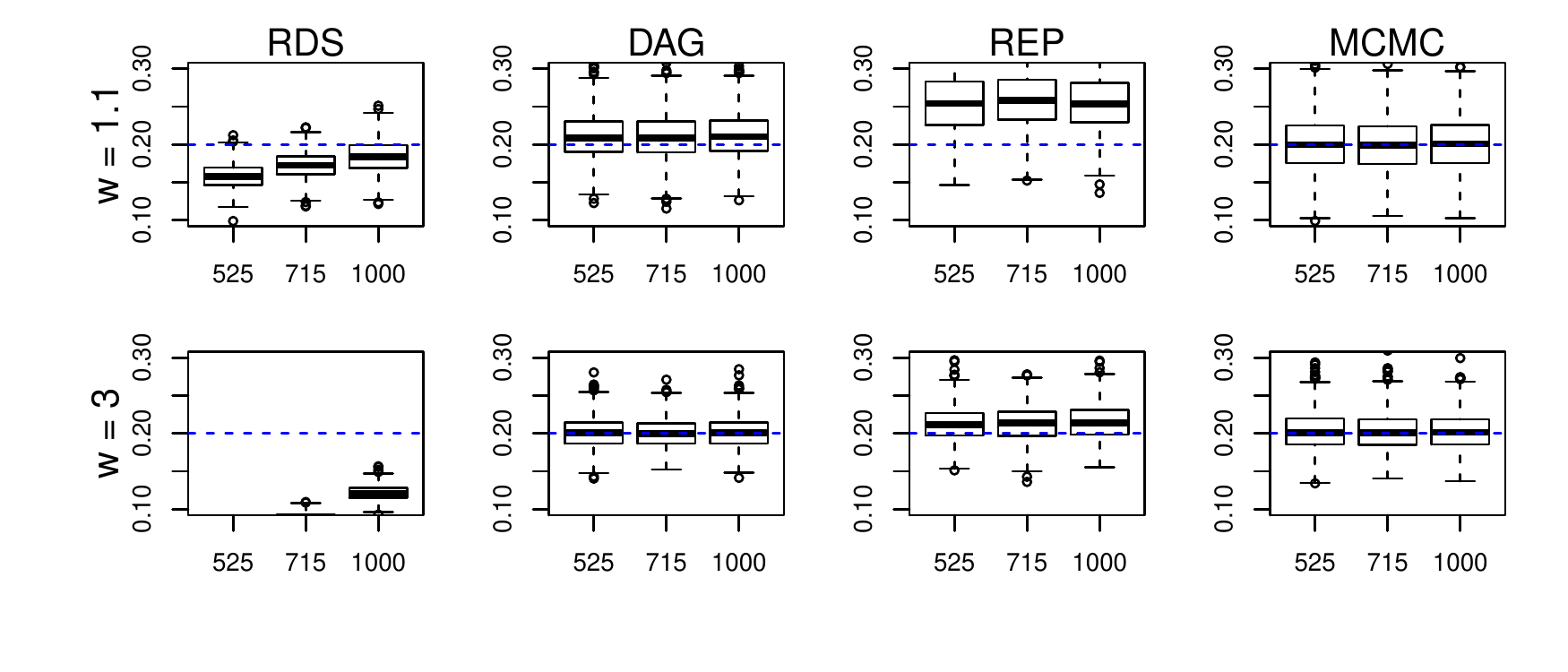}
\vspace{-1cm}
\caption{\label{fig:rdsrep-inf-burn} Estimated size of infected population
  where seeds are drawn from the infected population only with the first 100
nodes of each sample are discarded as ``burn-in.''}
\end{figure}

\begin{figure}
  \centering
  \includegraphics[scale=.70]{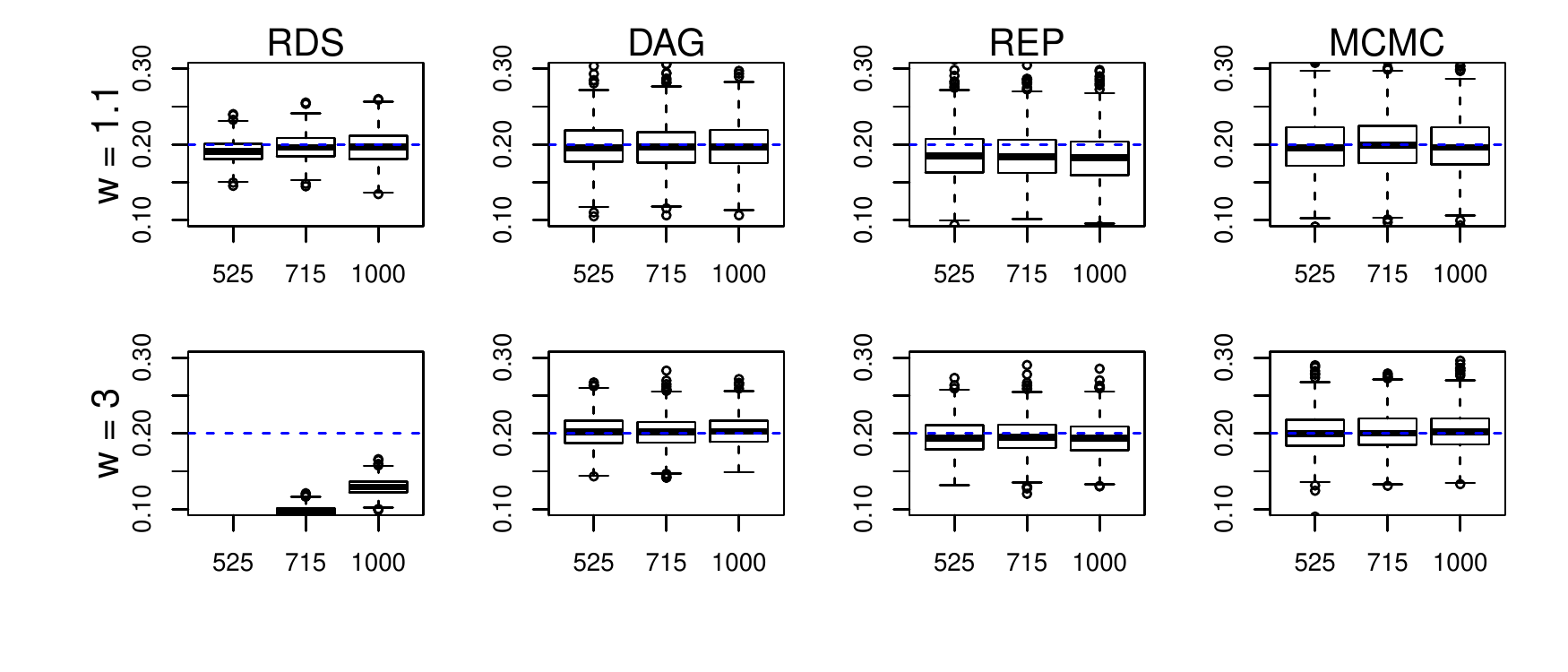}
\vspace{-1cm}
\caption{\label{fig:rdsrep-non-burn}Estimated size of infected population
  where seeds are drawn from the noninfected population only with the first 100
nodes of each sample are discarded as ``burn-in.''}
\end{figure}

Figures \ref{fig:rdsrep-no-bias}--\ref{fig:rdsrep-non-burn} show some
of our results. Here we run a series of tests, analogous to those Gile
and Handcock~\cite{gil-han:j:assess}.  All tests shown used a seed
size of six. The first three figures show
the effects of drawing seeds from the entire population, just the
infected population, and just the uninfected population, respectively.
Together, they show the effects of recruitment bias on the performance
of the estimators. 

Additionally, we consider \emph{burn-in},
a feature of most MCMC-based sampling in which a fraction of the
earliest samples are dropped, because they more heavily depend on the 
seeds---and are thus more biased---than the later samples, which
are ideally independent of the seeds. The last two figures show the
effects of recruitment bias after a burn-in of the first 100 samples.

The parameters considered within each figure are the network sizes 1000, 
715, and 525 and the activity ratios (labeled ``w'') 1.1 and 3. 


\section{Discussion and Conclusion}
The results for RDS and REP essentially
replicate for comparison purposes
those of Gile and Handcock. One reason RDS performance degrades so dramatically as 
network size decreases is that the probability that any node is sampled approaches
one as the network size decreases, but the VH estimator still weighs
each sample as if it had been chosen in proportion to its network
neighborhood.

Of all the protocols we test, MCMC performs best, which is what we would expect
as it represents RDS in the impractical case when all the VH
assumptions hold. Surprisingly to us, 
DAG was clearly second best, outperforming even REP, the protocol which seemed
to us to be the most like MCMC (note that both REP and MCMC produce
arbitrary recruitment networks). The only test in which DAG did not
perform at a level comparable to MCMC was when all seed nodes were
drawn from the infected population and the activity ratio was low,
though a 100 node burn-in almost corrects this.
We are investigating why DAG performs as well as it
has. Space prevents us from giving details, but we have seen that
the recruitment graphs created by DAG have clustering coefficients and
average path lengths that are closer than the other protocols to
MCMC. 

We hope that this study shows that creative thinking about how RDS is
implemented in OSNs may lead to significant improvements in its
sampling accuracy. We have ideas about how human-computer inferace
methods on OSNs can improve
neighborhood size recall and the randomness of the recruitment
process, neither of which we have space to discuss here. Additional
open issues remain, such as the inherent biases of OSNs and the
degree of realism that the ERGM models used here and in related work
provide. In future work we plan to conduct field studies of these issues
and others, using an actual implementation of RDS over Facebook.
\bibliographystyle{alpha}
\bibliography{harish.bibliography}

\newcommand{\etalchar}[1]{$^{#1}$}
\begin{thebibliography}{AQHM{\etalchar{+}}06}

\bibitem[AQHM{\etalchar{+}}06]{abdul2006effectiveness}
A.S. Abdul-Quader, D.D. Heckathorn, C.~McKnight, H.~Bramson, C.~Nemeth,
  K.~Sabin, K.~Gallagher, and D.C. Des~Jarlais.
\newblock Effectiveness of respondent-driven sampling for recruiting drug users
  in {N}ew {Y}ork {C}ity: findings from a pilot study.
\newblock {\em Journal of Urban Health}, 83(3):459--476, 2006.

\bibitem[BSPar]{bernhardt2012all}
A.~Bernhardt, M.~Spiller, and D.~Polson.
\newblock All work and no pay: Violations of employment and labor laws in
  {C}hicago, {L}os {A}ngeles and {N}ew {Y}ork {C}ity.
\newblock {\em Social Forces}, to appear.

\bibitem[GH10]{gil-han:j:assess}
K.~Gile and M.~Handcock.
\newblock Respondent-driven sampling: an assessment of current methodology.
\newblock {\em Sociological Methodology}, 40(1):285--327, August 2010.

\bibitem[GH11]{gile2011network}
K.J. Gile and M.S. Handcock.
\newblock Network model-assisted inference from respondent-driven sampling
  data.
\newblock {\em arXiv preprint arXiv:1108.0298}, 2011.

\bibitem[Gil11]{gile2011improved}
K.J. Gile.
\newblock Improved inference for respondent-driven sampling data with
  application to {HIV} prevalence estimation.
\newblock {\em Journal of the American Statistical Association},
  106(493):135--146, 2011.

\bibitem[GJS12]{gile2012diagnostics}
K.J. Gile, L.G. Johnston, and M.J. Salganik.
\newblock Diagnostics for respondent-driven sampling.
\newblock {\em arXiv preprint arXiv:1209.6254}, 2012.

\bibitem[GKBM11]{Gjoka2011}
M.~Gjoka, M.~Kurant, C.~Butts, and A.~Markopoulou.
\newblock A walk in facebook: Uniform sampling of users in online social
  networks.
\newblock Technical Report 0906.0060v4, arXiv, February 2011.

\bibitem[GS10]{goel_salganik_2010}
Sharad Goel and Matthew~J Salganik.
\newblock Assessing respondent-driven sampling.
\newblock {\em Proceedings of the National Academy of Sciences of the United
  States of America}, 107(1515):6743--6747, 2010.

\bibitem[Hec97]{hec:j:rdsone}
D.~Heckathorn.
\newblock Respondent-driven sampling: A new approach to the study of hidden
  populations.
\newblock {\em Social Problems}, 44(2):174--199, May 1997.

\bibitem[Hec07]{hec:u:rds}
D.~Heckathorn.
\newblock Extensions of respondent-driven sampling: Analyzing continuous
  variables and controlling for differential recruitment.
\newblock {\em Sociological Methodology}, 37(1):151--207, December 2007.
\newblock in press.

\bibitem[HG10]{handcock2010modeling}
M.S. Handcock and K.J. Gile.
\newblock Modeling social networks from sampled data.
\newblock {\em The Annals of Applied Statistics}, 4(1):5--25, 2010.

\bibitem[HJ01]{heckathorn2001finding}
D.~Heckathorn and J.~Jeffri.
\newblock Finding the beat: Using respondent-driven sampling to study jazz
  musicians.
\newblock {\em Poetics}, 28(4):307--329, 2001.

\bibitem[MJK{\etalchar{+}}08]{malekinejad2008using}
M.~Malekinejad, L.G. Johnston, C.~Kendall, L.R.F.S. Kerr, M.R. Rifkin, and G.W.
  Rutherford.
\newblock Using respondent-driven sampling methodology for {HIV} biological and
  behavioral surveillance in international settings: a systematic review.
\newblock {\em AIDS and Behavior}, 12:105--130, 2008.

\bibitem[SH04]{salganik2004sampling}
M.J. Salganik and D.D. Heckathorn.
\newblock Sampling and estimation in hidden populations using respondent-driven
  sampling.
\newblock {\em Sociological methodology}, 34(1):193--240, 2004.

\bibitem[TG11]{tomas2011effect}
A.~Tomas and K.J. Gile.
\newblock The effect of differential recruitment, non-response and
  non-recruitment on estimators for respondent-driven sampling.
\newblock {\em Electronic Journal of Statistics}, 5:899--934, 2011.

\bibitem[Tho92]{tho:b:sampling}
S.~Thompson.
\newblock {\em Sampling}.
\newblock John Wiley \& Sons, New York NY, 1992.

\bibitem[VH08]{hec:j:rds-theory-08}
E.~Volz and D.~Heckathorn.
\newblock Probability based estimation theory for respondent driven sampling.
\newblock {\em Journal of Official Statistics}, 24(1):79--97, 2008.

\bibitem[Wej09]{wejnert_2009}
C~Wejnert.
\newblock An empirical test of respondent-driven sampling: Point estimates,
  variance, degree measures, and out-of-equilibrium data.
\newblock {\em Sociological Methodology}, 39(1):73--116, August 2009.

\bibitem[WH08]{wejnert_heckathorn_2008}
C.~Wejnert and D.~Heckathorn.
\newblock Web-based network sampling: Efficiency and efficacy of
  respondent-driven sampling for online research.
\newblock {\em Sociological Methods \& Research}, 37(1):105--134, June 2008.

\bibitem[WS02]{walters2002health}
K.~Walters and J.~Simoni.
\newblock Health survey of two-spirited native americans, 2002.

\end{thebibliography}
\end{document}